\documentclass[11pt]{article}
\usepackage{amsfonts}
\usepackage{amsmath}
\usepackage{amssymb}
\usepackage{latexsym}
\usepackage{times}
\usepackage{epsfig}
\usepackage{multirow}
\usepackage{hyperref}
\usepackage{url}
\newcommand{\ket}[1]{|\, #1\rangle}
\newcommand{\bra}[1]{\langle #1\,|}

\newcommand\beq{\begin{equation}}
\newcommand\eeq{\end{equation}}
\newcommand\bea{\begin{eqnarray}}
\newcommand\eea{\end{eqnarray}}

\renewcommand{\det}[1]{\mbox{\sf det}\left(#1\right)} 

 \def\squarebox#1{\hbox
to #1{\hfill\vbox to #1{\vfill}}}
\def\qed{\hspace*{\fill}\vbox{\hrule\hbox{\vrule\squarebox{.667em}\vrule}\hrule}
} 
\newenvironment{proof}{\begin{trivlist}\item[]{\bf Proof:}}{\qed
\end{trivlist}}

\newcommand{\ba}{\begin{array}}
\newcommand{\ea}{\end{array}}

\newtheorem{theo}{Theorem}

\newtheorem{lem}{Lemma}

\newtheorem{cor}{Corollary}

\begin{document}

\author{David P. DiVincenzo and Barbara M. Terhal\footnote{IBM Watson Research Center, P.O. Box 218, Yorktown
Heights, NY 10598, USA. }  }

\title{Fermionic Linear Optics Revisited}

\date{\today}

\maketitle

\begin{abstract}
We provide an alternative view of the efficient classical
simulatibility of fermionic linear optics in terms of Slater
determinants. We investigate the generic effects of two-mode
measurements on the Slater number of fermionic states.  We argue
that most such measurements are not capable (in conjunction with
fermion linear optics) of an efficient {\bf exact} implementation of universal quantum
computation. Our arguments do not apply to the two-mode parity measurement, for
which exact quantum computation becomes possible, see \cite{beenakker+:ferm}.
\end{abstract}

\subsection*{Dedication to Asher Peres}

It is a pleasure to contribute to the Festschrift for Asher
Peres' 70th birthday. To characterize Asher's achievements in
physics and quantum information theory, we would like to quote
from a novel by the Austrian writer Robert Musil (1880-1942) {\em
The man without qualities} written in 1930 \cite{book:musil}. The
protagonist of the novel, Ulrich, is full of praise about science:

\begin{quotation}
But one thing, on the other hand, could safely be said about
Ulrich: he loved mathematics because of the kind of people who
could not endure it. He was in love with science not so much on
scientific as on human grounds. He saw that in all the problems
that come within its orbit, science thinks differently from the
laity. If we translate ``scientific outlook" into ``view of life,"
``hypothesis" into ``attempt," and ``truth" into ``action," then
there would be no notable scientist or mathematician whose life's
work, in courage and revolutionary impact, did not far outmatch
the greatest deeds of history. The man has not yet been born who
could say to his followers: "You may steal, kill, fornicate - our
teaching is so strong that it will transform the cesspool of your
sins into clear, sparkling mountain streams." But in science it
happens every few years that something till then held to be in
error suddenly revolutionizes the field, or that some dim and
disdained idea becomes the ruler of a new realm of thought. Such
events are not merely upheavals but lead us upward like a Jacob's
ladder. The life of science is as strong and carefree and glorious
as a fairy tale. And Ulrich felt: People simply don't realize it,
they have no idea how much thinking can be done already; if they
could be taught to think a new way, they would change their lives.
\end{quotation}

Asher Peres is one of those scientists whose work exemplifies the
force of logical thinking and of independent and unorthodox
investigation into the nature of physics, as demonstrated by
his wide ranging publications from relativity theory to quantum
mechanics. We are grateful for his ``revolutionary and courageous acts" in
quantum information theory which have been an inspiration for our
own work. We also hope that Asher will appreciate this small paper
whose subject is at the boundary of physics and information, a
boundary that Asher has enjoyed crossing during his productive
scientific career.

\section{Fermionic linear optics and single-mode measurements}
\label{preserve}

A short introduction to second quantization \footnote{Apparently,
the second-quantized analysis of fermions developed over the
 course of a series of papers, principally by P. Jordan; the most
notable in this series is P. Jordan and E. P. Wigner, ``About the
Pauli exclusion principle", Z. Physik {\bf 47}, 631 (1928).  It
had assumed essentially its modern form by the time of V. A. Fock,
``Configuration space and second quantization", Z. Physik {\bf
75}, 622 (1932). It is reviewed  in innumerable modern texts,
e.g., A. L. Fetter and J. D. Walecka, {\it Quantum theory of
many-particle systems}, (McGraw-Hill, New York, 1971), Chap. 1.}
will serve to set our notation.  Suppose that the Hilbert space of
a single electron has dimension $D$, and is spanned by a standard
basis $\ket{i}$, $1\leq i\leq D$.  The label $i$ may indicate both
spin and space degrees of freedom.  In the language of second
quantization these same basis vectors are indicated as
$a_i^\dagger\ket{\bf 0}$.  $\ket{\bf 0}$ is the basis vector of
the Hilbert space corresponding to no electrons (the {\it vacuum}
state). $a_i^\dagger$ is the {\it creation operator for an
electron in mode (or orbital) $i$}. Without the dagger it is a
{\it destruction operator}. Creation operators are taken to
anticommute \footnote{$\{a_i^{\dagger},a_j^{\dagger}\}=\{a_i,a_j\}=0$ and $\{a_i,a_j^{\dagger}\}=\delta_{ij}I$.}
which enforces the Pauli principle since
$(a_i^\dagger)^2=0$, i.e., two electrons cannot be put in the same
state.

We are interested in fermionic states that contain not just one
electron but $N\leq D$ electrons.  In second quantized language,
an example of such a state is
\begin{equation}
\Pi_{i=1}^Na_i^\dagger\ket{\bf 0}. \label{singledet}
\end{equation}
This state is special not because it places electrons in the
standard-basis orbitals, but because it can be written as a single
term.  In the old language of electron physics, Eq.
(\ref{singledet}) is an example of a single Slater determinant.  A
general $N$-electron state is a superposition of such Slater
determinants, in which the product over orbitals can run over any
set of $N$ {\em orthogonal} single-particle basis states.

Recent work on quantum information processing with fermions has
focussed on this Slater-determinant characterization of electron
states.  For example, a proposed measure of entanglement of fermi
systems \cite{schliemann+:slater1,eckert+:slater2} is the ``Slater
number", the number of terms in the expression for the wave
function involving the least number of Slater determinants.

It has been known since the very earliest work in computational
physics that the simulation of physical properties of electron
systems becomes tractable if the states can be approximated by
single Slater determinants (this is an essential feature of the
Hartree-Fock approximation).  This approximation is exact if the
Coulomb interaction between electrons can be ignored.  In fact
this is a rather drastic approximation, and much of the art of
atomic (and solid state) modelling has consisted of finding well
chosen ``mean fields", atomic potentials that mimic as well as
possible the average effect of the interaction of many electrons
in the atom. This endeavor has been rather successful, and has
provided a basis for the approximate computation of many quantum
properties in atomic, molecular, and solid-state physics, as well
as in chemistry.

Recent work by the authors \cite{TD:fermions} (see also Ref.
\cite{knill:flo}) has shown that the restricted quantum
computational process of fermionic linear optics can be simulated
efficiently on a classical computer. Fermionic linear optics on a
set of non-interacting electrons are operations such as beam
splitters, phase-shifters (delay lines), von Neumann measurements
of the electron state, with the choice of quantum operations potentially
based on prior measurement results.

It was not explicitly shown in \cite{TD:fermions} that the Slater
number remains one under these operations, and we will show it
here and argue that, perhaps not surprisingly, it provides the
basis for the classical simulatibility. First, the allowed
Hamiltonian evolutions in this model are in the class of
``one-body interactions"; that is, they arise from forces between
the electrons and the controlling apparatus, and not between
different electrons.  Such Hamiltonians $H_1$ have non-zero matrix
elements only between $N$-particle Slater determinant states
$\Phi_1$ and $\Phi_2$ with the same $N$, i.e., the Hamiltonian is
number conserving (although an extension to fermion-parity
conserving Hamiltonians is possible, and has been worked out in
\cite{TD:fermions, knill:flo}); $N-1$ of the orbitals in $\Phi_1$
and $\Phi_2$ should be identical, and just one may be different in
the two states. For example, generally
\begin{equation}
\bra{\Phi_1}H_1\ket{\Phi_2}\neq 0\,\,\, \mbox{if}
\,\,\ket{\Phi_1}=\Pi_{i=1}^Na_i^\dagger\ket{{\bf
0}},\,\ket{\Phi_2}=(\Pi_{i=1}^{N-1}a_i^\dagger)a_{N+1}^\dagger\ket{\bf
0}
\end{equation}
but
\begin{equation}
\bra{\Phi_1}H_1\ket{\Phi_3}= 0\,\,\, \mbox{if}
\,\,\ket{\Phi_1}=\Pi_{i=1}^Na_i^\dagger\ket{{\bf
0}},\,\ket{\Phi_3}=(\Pi_{i=1}^{N-2}a_i^\dagger)a_{N+1}^\dagger
a_{N+2}^\dagger\ket{\bf 0}
\end{equation}
In this last example, the matrix element would be nonzero if two
body terms in the Hamiltonian (electron-electron interactions)
were included. In general, such a non-interacting Hamiltonian can
be written as

\begin{equation}
H(t)=\sum_{i,j=1}^D b_{ij}(t) a_i^{\dagger}a_j.\label{aexpr}
\end{equation}
Equation (\ref{aexpr}) introduces the Hermitian time-dependent
matrix {\bf b}(t).  We will use the standard result, reviewed in
\cite{TD:fermions}, that the action of the time evolution
operator,
\begin{equation}
U(\tau)=T\exp(-i\int_0^\tau dtH(t))
\end{equation}
($T$ is the time-ordering operator), on a creation operator is
\begin{equation}
U(\tau)a_i^\dagger U^\dagger(\tau)=\sum_m
V_{im}(\tau)a_m^\dagger=a_i^{\dagger}(\tau).\label{timedev}
\end{equation}
Here the unitary matrix {\bf V} is given by
\begin{equation}
{\bf V}(\tau)=T\exp(-i\int_0^\tau dt\,{\bf b}(t)).
\end{equation}
The notation introduced in the last part of Eq. (\ref{timedev})
(the $\tau$ dependence) indicates that the resulting operator is
just the creation operator for an electron in the time-evolved
orbital
\begin{equation}
\ket{i(\tau)}=\sum_m V_{im}(\tau)\ket{m}.
\end{equation}
Under $U(\tau)$, then, the initial state Eq. (\ref{singledet})
evolves to (using $U \ket{\bf 0}=\ket{{\bf 0}}$),
\begin{equation}
\Pi_{i=1}^N a_i^{\dagger}(\tau)\ket{\bf 0}, \label{singledetstill}
\end{equation}
i.e., still a single Slater determinant in a rotated basis.

We now turn to the other computational step considered by
\cite{TD:fermions}, projective measurement of the occupation of a
single orbital (call it $\ket{\kappa}$).  The projector
corresponding to the state being occupied is
\begin{equation}
\textsf{P}_1=a_\kappa^\dagger a_\kappa,
\end{equation}
and for the unoccupied outcome, the projector is
\begin{equation}
\textsf{P}_0=1-a_\kappa^\dagger a_\kappa=a_\kappa
a_\kappa^\dagger.
\end{equation}
What is noteworthy is that both projectors consist of a single product of annihilation and
creation operators, which would not be the case for bosons.

Let us now show that the state after measurement, under all
circumstances, continues to be a single Slater determinant.  That
is, we show that
\begin{equation}
\textsf{P}_0 (\Pi_{i=1}^N a_i^{\dagger}(\tau))\ket{\bf 0}
\label{singledetstill2}
\end{equation}
and
\begin{equation}
\textsf{P}_1 (\Pi_{i=1}^Na_i^{\dagger}(\tau))\ket{\bf 0}
\label{singledetstill3}
\end{equation}
are single Slater determinants, of a very simple form.  The
technical steps are described in an Appendix of
\cite{eckert+:slater2}, we give a simple version of them here for
completeness.  We first write the orbital $\ket{\kappa}$ as a
linear combination of an orbital in the space spanned by the set
$\{\ket{i(\tau)}\}$, and an orbital not in that span:
\begin{equation}
\ket{\kappa}=\alpha\ket{\mbox{in}}+\beta\ket{\mbox{out}}
\end{equation}
where the two new normalized vectors are defined by
\begin{equation}
\ket{\mbox{in}}\in\mbox{Span}(\{\ket{i(\tau)}\}),
\end{equation}
\begin{equation}
\ket{\mbox{out}}\in\mbox{Ker}(\{\ket{i(\tau)}\}).
\end{equation}
The phase of these orbitals can be chosen so that the coefficients
$\alpha$ and $\beta$ are real and nonnegative, and
$\alpha^2+\beta^2=1$.  We also rewrite the Slater determinant in
terms of a new orthogonal basis of orbitals:
\begin{equation}
\Pi_{i=1}^N a_i^{\dagger}(\tau)\ket{{\bf
0}}=a_{\mbox{in}}^\dagger\Pi_\mu a_\mu^\dagger\ket{\bf
0}.\label{rotate}
\end{equation}
Here we introduce new orbitals $\ket{mu}$, $\mu=2,3,...,N$, such
that
$\mbox{Span}(\{\ket{i(\tau)}\})=\mbox{Span}(\ket{\mbox{in}},\{\ket{\mu}\})$,
that is, the space of filled states remains the same.  Eq.
(\ref{rotate}) can be shown by the following steps. Let $U$ be a
fermionic linear-optics transformation such that for $i=1,\ldots
N$
\begin{equation}
U a_i^\dagger U^\dagger=\sum_{m=1}^D V_{im} a_m^\dagger,
\end{equation}
where the $D \times D$ unitary matrix $V={\cal V} \oplus {\cal W}$
is a block-diagonal matrix with ${\cal V} \in {\rm SU}(N)$.
Furthermore ${\cal V}$ is such that $\ket{1(\tau)}$ is rotated to
$\ket{\mbox{in}}$ and $\ket{i(\tau)}$, $i>1$, is rotated to
$\ket{\mu}$, $\mu=2,3,...,N$. Thus, as required, the rotation
${\cal V}$ does not change the space of filled orbitals. By
inserting $U^{\dagger}U={\bf I}$ and using $U \ket{\bf 0}=\ket{\bf
0}$ we can see that \beq U\, \Pi_{i=1}^N
a_i^{\dagger}(\tau)\ket{{\bf 0}}=a_{\mbox{in}}^\dagger\Pi_\mu
a_\mu^\dagger\ket{\bf 0}, \eeq while we can also write
\begin{equation}
U\,\,\Pi_{i=1}^N a_i^\dagger=\sum_{i=1}^N
{\cal V}_{1i}a_i^\dagger\sum_{j=1}^N {\cal V} _{2j}a_j^\dagger\sum_{k=1}^N
{\cal V}_{3k}a_k^\dagger...
\end{equation}
There are $N$ modes, and $N$ sums in this expression.  In order
for a term to be nonzero, each mode must appear once and only once
(otherwise a mode would be repeated, and $(a^\dagger)^2=0$).  Thus
we can write using the anti-commutation relations that
\begin{equation}
U\,\,\Pi_{i=1}^N a_i^\dagger=\sum_\pi\mbox{sign}(\pi){\cal V}_{1,\pi(1)}{\cal V}_{2,\pi(2)}
{\cal V}_{3,\pi(3)}
\Pi_{i=1}^Na_i^\dagger=\det {\cal V}\,\Pi_{i=1}^Na_i^\dagger=\Pi_{i=1}^Na_i^\dagger.
\end{equation}
Thus, Eq.(\ref{rotate}) is established.

Now we can calculate:
\begin{eqnarray}
\textsf{P}_1(\Pi_{i=1}^N a_i^{\dagger}(\tau)\ket{{\bf 0}}&=&
a_\kappa^\dagger a_\kappa(\Pi_{i=1}^N a_i^{\dagger}(\tau)\ket{{\bf
0}}\nonumber\\&=&a_\kappa^\dagger (\alpha a_{\mbox{in}}+\beta
a_{\mbox{out}})a_{\mbox{in}}^\dagger\Pi_\mu a_\mu^\dagger\ket{{\bf
0}}=\alpha a_\kappa^\dagger\Pi_\mu a_\mu^\dagger\ket{{\bf 0}}.
\end{eqnarray}
Here we have used the fact that $a_{\mbox{out}}\ket{{\bf 0}}=0$
and $a_{\mbox{in}}a_{\mbox{in}}^\dagger\ket{{\bf 0}}=\ket{{\bf
0}}$. We note that the final state is indeed just another Slater
determinant; it is unnormalized, but the coefficient $\alpha$ just
reflects the fact that the probability of this outcome is
$\alpha^2$.  Note that $\alpha$ is easy to calculate, as it is
just the magnitude of the projection of a vector in the
single-particle space of dimension $D$. The outcome of the other
projector just takes a little bit more to evaluate:
\begin{eqnarray}
\textsf{P}_0(\Pi_{i=1}^N(a_i^\tau)^\dagger)\ket{{\bf
0}}&=&a_\kappa a_\kappa^\dagger(\Pi_{i=1}^N
a_i^{\dagger}(\tau)\ket{{\bf 0}}\nonumber\\&=&a_\kappa (\alpha
a_{\mbox{in}}^\dagger+\beta
a_{\mbox{out}}^\dagger)a_{\mbox{in}}^\dagger\Pi_\mu
a_\mu^\dagger\ket{{\bf 0}}=\beta a_\kappa a_{\mbox{out}}^\dagger
a_{\mbox{in}}^\dagger\Pi_\mu a_\mu^\dagger\ket{{\bf 0}}.
\end{eqnarray}
Here we have used $(a_{\mbox{in}}^\dagger)^2=0$.  To go further,
we have to introduce another orbital $\ket{\kappa^\perp}$
perpendicular to $\ket{\kappa}$ in the space spanned by
$\ket{\mbox{in}}$ and $\ket{\mbox{out}}$; specifically,
\begin{equation}
\ket{\kappa^\perp}=\beta\ket{\mbox{in}}-\alpha\ket{\mbox{out}}.
\end{equation}
Then using the relation
\begin{equation}
a_{\mbox{out}}^\dagger a_{\mbox{in}}^\dagger=a_\kappa^\dagger
a_{\kappa^\perp}^\dagger,
\end{equation}
we finish the derivation:
\begin{eqnarray}
\textsf{P}_0(\Pi_{i=1}^N a_i^{\dagger}(\tau)\ket{{\bf 0}}=\beta
a_\kappa a_{\mbox{out}}^\dagger a_{\mbox{in}}^\dagger\Pi_\mu
a_\mu^\dagger\ket{{\bf 0}}=\beta a_\kappa a_\kappa^\dagger
a_{\kappa^\perp}^\dagger\Pi_\mu a_\mu^\dagger\ket{{\bf 0}}=\beta
a_{\kappa^\perp}^\dagger\Pi_\mu a_\mu^\dagger\ket{{\bf 0}}.
\end{eqnarray}

So all operations keep the state vector in the Slater-determinant
form.  Indeed, we note that except for the change of
normalization, the action of $\textsf{P}_{0,1}$ on the state is
identical to that of some single-particle Hamiltonian.  This gives
a new view on the classical simulatability of this suite of
operations. Appendix A gives some correspondence between these
calculations and ones that appear in the theory of electron energy
bands in crystals.

\section{Two-mode measurements}
\label{twomode}

We now consider a different scenario for quantum computation, one
in which one can perform a two-mode measurement.  This can for
example be a ``charge measurement" that determines how many
electrons are in a particular spatial orbital irrespective of
their spin state. This means that we imagine a two-mode
measurement, in which the two modes are identical spatially and
differ only in their spin quantum number (so we might indicate
these two orbitals $\ket{m\uparrow}$ and $\ket{m\downarrow}$, for
some spatial orbital $m$).  In the present analysis we will not
wish to make the distinction between spin and orbital labels, so
that we will just consider a number measurement in two orthogonal
modes labelled $\ket{\kappa}$ and $\ket{\lambda}$.  (This
generalization means that our analysis is applicable to problems
involving spin-orbit interaction, where the distinction between
spin and space is not applicable.)

We will assume that this measurement is ``nondestructive", a
feature of recent charge measurements in quantum dot structures
\cite{hanson+:qd} (but, see the remarks in the Discussion
section). Then, similar to a single-mode measurement, we must
write down the `operation elements' (see
\cite{book:nielsen&chuang}) of the measurement which we take to be
projectors (see the Discussion for potential modifications).

For the ``0" outcome (both modes unoccupied), this is simple,
since it is just given by the product of the two ``0" projectors
for the two modes separately:
\begin{equation}
\textbf{\textsf{P}}_0=\textsf{P}_{0\kappa}\textsf{P}_{0\lambda}=a_\kappa
a_\kappa^\dagger a_\lambda a_\lambda^\dagger.
\end{equation}
The ``2" outcome projector (both orbitals occupied) is also simply
the product of the two one-orbital projectors:
\begin{equation}
\textbf{\textsf{P}}_2=\textsf{P}_{1\kappa}\textsf{P}_{1\lambda}=a_\kappa^\dagger
a_\kappa a_\lambda^\dagger a_\lambda.
\end{equation}
The ``1" projector can be written as a {\em sum} of two projector
products:
\begin{equation}
\textbf{\textsf{P}}_1=a_\kappa a_\kappa^\dagger a_\lambda^\dagger
a_\lambda+a_\kappa^\dagger a_\kappa a_\lambda a_\lambda^\dagger.
\end{equation}

The important point for the upcoming analysis is

\begin{lem}
$\textbf{\textsf{P}}_1$ cannot be expressed as a single term; that
is, it is not possible to write $\textbf{\textsf{P}}_1$ as
$\textbf{\textsf{P}}_1=f_1f_2f_3...f_M$, where $M$ is an arbitrary
integer, and each $f_i$ is either a creation or an annihilation
operator for some arbitrary mode.
\end{lem}

\begin{proof} We have to study two cases:

1) $f_M$ is a creation operator.  We consider a one-particle basis
consisting of orbitals $\ket{\kappa}$ and $\ket{\lambda}$, and
$D-2$ orbitals $\ket{\nu}$ orthogonal to $\ket{\kappa}$ and
$\ket{\lambda}$.  Then the orbital $\phi_M$ that $f_M$ creates can
be written
\begin{equation}
\ket{\phi_M}=\alpha\ket{\kappa}+\beta\ket{\lambda}+\sum_{\nu}
c_\nu\ket{\nu}.\label{fexpand}
\end{equation}
Consider the (unnormalized) state
\begin{equation}
\ket{\Phi}=(\alpha a_\kappa^\dagger+\beta
a_\lambda^\dagger)\Pi_\nu a_\nu^\dagger\ket{{\bf 0}}.
\end{equation}
This state has one electron in the space ${\rm
Span}(\ket{\kappa},\ket{\lambda})$, so it is an eigenstate of
$\textbf{\textsf{P}}_1$ with eigenvalue 1:
\begin{equation}
\textbf{\textsf{P}}_1\ket{\Phi}=\ket{\Phi}.
\end{equation}
But $f_M$ annihilates $\ket{\Phi}$ (recall that
$(a_x^\dagger)^2=0$ for any $x$):
\begin{eqnarray}
f_M\ket{\Phi}=(\alpha a_\kappa^\dagger+\beta
a_\lambda^\dagger+\sum_{\nu'} c_{\nu'} a_{\nu'}^\dagger)(\alpha
a_\kappa^\dagger+\beta a_\lambda^\dagger)\Pi_\nu
a_\nu^\dagger\ket{{\bf 0}}\nonumber\\
=[(\alpha a_\kappa^\dagger+\beta a_\lambda^\dagger)^2\Pi_\nu
a_\nu^\dagger-\sum_{\nu'} c_{\nu'}(\alpha a_\kappa^\dagger+\beta
a_\lambda^\dagger)a_{\nu'}^\dagger\Pi_\nu a_\nu^\dagger]\ket{{\bf
0}}=0,
\end{eqnarray}
so the single-term expression $f_1f_2f_3...f_M$ cannot equal
$\textbf{\textsf{P}}_1$ in this case.

2) $f_M$ is an annihilation operator.  We consider the same
orbital expansion as in Eq. (\ref{fexpand}), and we apply $f_M$ to
the state
\begin{equation}
\ket{\Psi}=(\alpha^* a_\kappa+\beta^* a_\lambda)a_\kappa^\dagger
a_\lambda^\dagger\ket{{\bf 0}}
\end{equation}
This is again an eigenstate of $\textbf{\textsf{P}}_1$,
$\textbf{\textsf{P}}_1\ket{\Psi}=\ket{\Psi}$. However,
\begin{eqnarray}
f_M\ket{\Psi}=(\alpha^* a_\kappa+\beta^* a_\lambda+\sum_{\nu'}
c_{\nu'}^* a_{\nu'})(\alpha^* a_\kappa+\beta^*
a_\lambda)a_\kappa^\dagger
a_\lambda^\dagger\ket{{\bf 0}}\nonumber\\
=[(\alpha^* a_\kappa+\beta^* a_\lambda)^2a_\kappa^\dagger
a_\lambda^\dagger-\sum_{\nu'} c_{\nu'}^*(\alpha^* a_\kappa+\beta^*
a_\lambda)a_\kappa^\dagger a_\lambda^\dagger a_{\nu'}]\ket{{\bf
0}}=0
\end{eqnarray}
(since $(a_x)^2=0$ and $a_x\ket{{\bf 0}}=0$).  So, in this case
too the single-term expression cannot match
$\textbf{\textsf{P}}_1$.
\end{proof}

So, the fact that $\textbf{\textsf{P}}_1$ cannot be written as a
single term opens the possibility that this two-mode measurement
can lead to more complex quantum time evolution, and thus has the
possibility of implementing quantum computation.  In particular,
when the ``1" outcome is obtained, the fact that the minimal
expression for $\textbf{\textsf{P}}_1$ contains two terms means
that the Slater number (recall above, see \cite{eckert+:slater2}))
could double after every such measurement; so, if there are $M$
such ``1" outcomes, then the state may have an exponentially large
($2^M$) Slater number, a state for which expectation values are
likely to be very hard to calculate classically.

We do not know how to prove that the Slater number will in fact be
as large as $2^M$ although we will now prove that generically,
when $\textbf{\textsf{P}}_1$ is applied to a {\em single} Slater
determinant, the result has Slater number two. Nevertheless, the
expectation that the Slater number becomes high, and the evolution
becomes difficult to simulate, is vindicated by the discovery of
Beenakker {\em et al.} \cite{beenakker+:ferm} that quantum
computation is implementable by linear fermion optics and the
two-mode measurement!  We will return to more discussion of this
measurement after Sec. \ref{digress}.

\section{Slater number generically goes from one to two under $\textbf{\textsf{P}}_1$}
\label{digress}

We now show that if we apply the two-mode projector
$\textbf{\textsf{P}}_1$ to a single Slater determinant for $N\geq
2$ electrons,
\begin{equation}
\ket{\Psi}=\textbf{\textsf{P}}_1\Pi_{i=1}^Na_i^\dagger\ket{{\bf
0}}=(a_\kappa a_\kappa^\dagger a_\lambda^\dagger
a_\lambda+a_\kappa^\dagger a_\kappa a_\lambda
a_\lambda^\dagger)\Pi_{i=1}^Na_i^\dagger\ket{{\bf 0}},
\end{equation}
then the resulting state $\ket{\Psi}$ generically has Slater
number two.  Note that we can always choose a basis such that the
initial state has the standard form shown.  Furthermore, without
loss of generality, we can parameterize the orthogonal orbitals
$\ket{\kappa}$ and $\ket{\lambda}$ as
\begin{eqnarray}
\ket{\kappa}&=&\cos\theta\ket{1}+\sin\theta\ket{N+1},\\
\ket{\lambda}&=&\cos\phi(-\sin\theta\ket{1}+\cos\theta\ket{N+1})+
\sin\phi(\cos\xi\ket{2}+\sin\xi\ket{N+2}).
\end{eqnarray}
We can simplify the problem considerably by using a theorem of K.
Eckert {\it et al.} \cite{eckert+:slater2}, that the Slater number
cannot be increased by applying an annihilation operator to a
state. Since $\ket{\kappa}$ and $\ket{\lambda}$ do not involve
orbitals $\ket{3}$, $\ket{4}$, ... $\ket{N}$, we can annihilate
all of these and be left with a two-electron state:
\begin{equation}
\ket{\Psi'}=\Pi_{i=3}^Na_i\ket{\Psi}=(a_\kappa a_\kappa^\dagger
a_\lambda^\dagger a_\lambda+a_\kappa^\dagger a_\kappa a_\lambda
a_\lambda^\dagger)a_1^\dagger a_2^\dagger\ket{{\bf 0}}.
\end{equation}
Using the methods of Sec. \ref{preserve}, we can convert each term
of this expression into one involving just two creation operators.
After a lengthy calculation (using Mathematica) we find
\begin{equation}
\ket{\Psi'}=\sum_{i,j=1,2,N+1,N+2}w_{ij}a_i^\dagger
a_j^\dagger\ket{{\bf 0}}.
\end{equation}
Where the $4\times 4$ antisymmetric matrix {\bf w} is

\begin{equation}
\left( \begin{array}{cccc} 0&{f_S} \,
     \cos \theta -
    {f_C} \,\sin \theta &
   -  \frac{\,\cos \xi \,
        \sin 2\phi}{{2f_S}}    &
   -  \frac{\cos \theta \,
          \sin^2\phi \,\sin 2\xi}{{2f_S}}    -
    \frac{\sin \theta \, \sin^2\phi \,
       \sin 2\xi}{{2f_C}}\\ \cdot& 0&
   -  {f_C} \,
       \cos \theta    -
    {f_S} \,\sin \theta &
   -  \frac{
        \sin 2\phi \,\sin \xi}{{2f_C}}\\ \cdot& \cdot
     &0&
   \frac{\cos \theta \,
        \sin^2\phi \,\sin 2\xi}{{
         2f_C}} -
    \frac{\sin \theta \, \sin^2\phi \,
     \sin 2\xi}{{2f_S}}\\ \cdot&\cdot&\cdot&0
\end{array}
\right)
\end{equation}
(we don't show the lower triangle of this antisymmetric matrix),
with
\begin{eqnarray}
f_C=\sqrt{\cos^2\phi + \cos^2\xi\sin^2\phi},\nonumber\\
f_S=\sqrt{\cos^2\phi + \sin^2\xi\sin^2\phi}.
\end{eqnarray}
As discussed by \cite{eckert+:slater2}, a basis transformation can
be made to bring an antisymmetric matrix to a canonical form,
consisting of a direct sum of $2\times 2$ antisymmetric blocks. The
number of nonzero blocks is the Slater number.  Obviously, for a
$4\times 4$ matrix the Slater number is two iff both blocks are
nonzero, and iff the determinant of the matrix is nonzero.  For an
antisymmetric matrix it is more convenient to evaluate the
Pfaffian, which is the square root of the determinant.  For {\bf
w} we find that the Pfaffian is
\begin{equation}
\mbox{Pf}({\bf w})={\sin^2\phi\sin 2\xi\over 2f_Sf_C}.
\end{equation}
So, we see that generically, $\textbf{\textsf{P}}_1$ does indeed
increase the Slater number from one to two.

\section{A no-go theorem}

We now explore further the power of nondestructive two-mode
measurements. As noted above, Beenakker {\em et al.}
\cite{beenakker+:ferm} have shown that the two-mode electric
charge measurement above, in conjunction with linear fermion
optics, permits the efficient implementation of quantum
computation.  It was noted, however, that like most of the linear
photon optics schemes proposed to date (cf. \cite{KLM:lo}), this
implementation using the three-outcome charge measurement is
non-deterministic, i.e., there is some finite chance that the
computation fails (in the case of an unlucky combination of
measurement outcomes), although the overall probability of failure
can be made acceptably low by a suitable implementation strategy.
We now argue that this small probability of failure is intrinsic
to this implementation:

\begin{theo}
If there exists an efficient implementation of the unitary
evolution of a quantum circuit using linear fermion optics
(including single-mode measurements) and the three-outcome,
two-mode measurement of Sec. \ref{twomode} that is {\bf exact},
i.e., has zero probability of failure, then this unitary evolution
has an efficient classical simulation. \label{theo1}
\end{theo}

\begin{proof} Suppose the exact implementation exists.
The efficient classical simulation of this unitary evolution
proceeds as follows: We begin with the standard, single Slater
determinant state of Eq. (\ref{singledet}). We compute the effect
of the first stage of fermionic linear optics on this state as in
Eq. (\ref{singledetstill}). Then, we consider the first 0/1/2
charge measurement. We can calculate whether the probability for
outcome ``1" is 100\% or not by computing the action of
$\textbf{\textsf{P}}_1$ on the state (a simple calculation for a
single Slater determinant).  If the probability of ``1" is 100\%,
if the state is an eigenstate of $\textbf{\textsf{P}}_1$ with
eigenvalue one, then the measurement has no effect on the state,
and we proceed on with the next stage of computation.  If the
probability of ``1" is not 100\%, then at least one of the
outcomes ``0" or ``2" has nonzero probability. Since the
implementation of the quantum gates is supposed to be exact (i.e.
works for every measurement outcome) we are free to choose any
outcome that occurs with non-zero probability. So we choose $0$ or
$2$ (making sure the choice has nonzero probability) and then note
that the state after application of $\textbf{\textsf{P}}_0$ or
$\textbf{\textsf{P}}_2$ is still a single Slater determinant. By
proceeding thus, the classical simulation at all stages need only
keep track of a single Slater determinant, which is efficiently
doable.
\end{proof}

{\em Remark}: In this proof we have assumed exact classical real-number
computation.  The proof can be relaxed to treat the case of finite
precision classical computations; in that case the quantum
computation is simulated approximately, but always with high
precision.

Thus if we believe (which we do) that there does not exist an
efficient classical simulation of the unitary evolution of
polynomially-sized quantum circuits \footnote{Note this does not
include the final single qubit measurements.}, it follows by this
Theorem that there will be no {\em exact} implementation of quantum
circuits using fermionic linear optics and the two-mode
three-outcome measurement.

We can modify the two-mode measurement such that some of the
outcomes are not distinguished and see what happens. For example,
we can consider a two-outcome measurement with projectors
$\textbf{\textsf{P}}_{0,1}=\textbf{\textsf{P}}_0+\textbf{\textsf{P}}_1$
and $\textbf{\textsf{P}}_2$, which only distinguishes whether the
two modes are completely filled or not. All such grouped
measurements can function in the nondeterministic implementation
of quantum computation of \cite{beenakker+:ferm}.  But

\begin{cor}
Theorem \ref{theo1} still holds if the three-outcome measurement
is replaced by the two-outcome measurement
$\textbf{\textsf{P}}_{0,1}/\textbf{\textsf{P}}_2$, or
$\textbf{\textsf{P}}_{1,2}/\textbf{\textsf{P}}_0$.
\end{cor}

\begin{proof} For both measurements there is an outcome ($\textbf{\textsf{P}}_2$ in the
first case, $\textbf{\textsf{P}}_0$ in the second) for which the
simulated state remains a single Slater determinant.  The rest of
the proof then applies.
\end{proof}

However, for one measurement (the ``parity" measurement), this
argument does not apply:

\smallskip
{\em The no-go theorem does not apply to the parity measurement
$\textbf{\textsf{P}}_{0,2}/\textbf{\textsf{P}}_1$.}
\smallskip

It would apply if one of the projectors could be written as a
single term.  We have already demonstrated that
$\textbf{\textsf{P}}_1$ cannot be written as a single term.  This
is easy to show for the projector
$\textbf{\textsf{P}}_{0,2}=\textbf{\textsf{P}}_0+\textbf{\textsf{P}}_2$
as well, by similar arguments: If $f_M$ (see Lemma 1) is a
creation operator, it annihilates the $D$-electron Slater
determinant, which is not annihilated by
$\textbf{\textsf{P}}_{0,2}$; if $f_M$ is a destruction operator,
it annihilates the vacuum $\ket{\bf 0}$, which is not annihilated
by $\textbf{\textsf{P}}_{0,2}$.

It was this observation that led Beenakker {\em et al.} to
investigate alternative implementations of quantum circuits using
linear fermion optics using the two-mode parity measurement; and,
indeed, an exact simulation, which is in some sense much more
efficient than any of the known non-deterministic simulations,
exists!

\section{Discussion}

We have presented an alternative description of the fermionic linear optics computation.
It is likely that the extension to ``fermion-parity preserving'' quadratic Hamiltonians
which was treated in Ref. \cite{TD:fermions}, can be analyzed similarly using Slater determinants.

We want to close with a few words of caution about the
applicability of our results.  We have indicated that the two-mode
measurement that enables quantum computation is ``nondestructive"
and uses projective measurement `elements'.  What happens if we relax these
conditions?

If the measurement is destructive, it means that the modes
$\ket{\kappa}$ and $\ket{\lambda}$ are no longer available for
further processing.  The ``tracing out" of these two modes that
this throwing away implies is implemented in second quantization
in the following way: the density matrix of the system, after the
application of the two-mode projectors discussed above, is changed
by the application of two trace-preserving completely positive
maps, ${\cal T_\kappa}$ and ${\cal T_\lambda}$. The
trace-over-$\zeta$ map ${\cal T_\zeta}$ is given
by\footnote{Actually, the simpler choice of Kraus operators
$A_1=a_\zeta^\dagger$, $A_2=a_\zeta$ has the equivalent effect.
This corresponds to going to a hole representation for the mode
$\zeta$.}
\begin{equation}
\rho'={\cal T}_\zeta(\rho)=\sum_{i=1}^2 A_i\rho
A_i^\dagger,\,\,A_1=a_\zeta
a_\zeta^\dagger,\,\,A_2=a_\zeta^\dagger a_\zeta.
\end{equation}
This map leaves the one-mode measurements
unchanged; but the two-mode measurements are changed in a very
important way. In particular, for any of the two-mode measurements
discussed above, the tracing out leaves the system in a density
matrix that is a mixture of single Slater determinants.  The
evolution of such states is simple (that is, efficiently
simulatable on a classical computer), so destructive measurements
give none of the quantum computational power of nondestructive
ones.  This has been anticipated in earlier studies of quantum
measurements for quantum computation
\cite{nielsen:comp,leung:comp}.

Another important modification of the measurement is the
following. Instead of the two-mode parity measurement with
measurement elements \cite{book:nielsen&chuang}
$\textbf{\textsf{P}}_{0,2}$ and $\textbf{\textsf{P}}_1$, suppose
we have a two-mode parity measurement with measurement elements
$U_{int}\textbf{\textsf{P}}_{0,2}$, (i.e. {\em not} a projector), and
$\textbf{\textsf{P}}_1$, where $U_{int}$ is a charge-preserving
unitary interaction. The probabilities of outcome of these two
measurements are the same but the state after the outcome $0/2$
has occurred will have undergone an additional unitary
transformation $U_{int}$ in the second type of measurement.

The status of the no-go theorems is the same for these two
measurements, but the construction in Ref.
\cite{beenakker+:ferm} only applies to the first.

This could be important, as there may be situations where
$U_{int}$ is nontrivial.  In particular, the Beenakker
construction \cite{beenakker+:ferm} is isomorphic to one in which
the qubit is coded by one electron in a double quantum dot, with
occupation of the orbital in the left dot representing $\ket{0}$
and right-dot orbital occupied being $\ket{1}$
\cite{barenco+:cond}. Then, as is suggested in
\cite{beenakker+:ferm}, the required charge parity measurement
might be accomplished by placing a single-electron transistor
between the right dot of one qubit and the left dot of the
adjacent qubit, so that it is sensitive to the charge on both (and
can be tuned so that it reads one value of current for both dots
empty or occupied, and another level otherwise).  However, an
analysis of this setup \cite{gurv} might reveal that $U_{int}$ is
nontrivial in this case due to an effective interaction which the
measurement sets up between the electrons in the two qubits.  More
analysis would be worthwhile in this case; and if, in this or in
similar situations, $U_{int}$ turns out to be different from the
identity, it would be worthwhile to work out an implementation of
quantum gates for this case.

\section*{Acknowledgments}

We are grateful for the support of the National Security Agency
and the Advanced Research and Development Activity through
contract DAAD19-01-C-0056.

\section*{Appendix A: Application to band theory}

The analysis of Sec. \ref{preserve} can be applied to simple model
problems in electron band theory.  Not surprisingly, given the
long history of band theory, see e.g. \cite{book:ashcroft&mermin},
some of the objects obtained above have special names, and special
significance, in this setting.

Suppose we consider a non-interacting Hamiltonian for electrons on
a one-dimensional lattice.  If the Hamiltonian only contains
nearest-neighbor hopping terms, $ta_i^\dagger a_{i+1}$, and $t<0$,
then the ground state of the system is the Slater determinant
\begin{equation}
\Pi_{|k|\leq k_F}^Na_k^\dagger\ket{\bf 0}. \label{fermisea}
\end{equation}
The orbital $a_k^{\dagger}\ket{{\bf 0}}=\ket{k}$ is the plane wave
\begin{equation}
\ket{k}={1\over\sqrt{D}}\sum_x e^{ikx}\ket{x}.
\end{equation}
Here $D$ is the number of lattice sites (we assume periodic
boundary conditions), $a_x^{\dagger}\ket{{\bf 0}}=\ket{x}$ is the
orbital centered at site $x$, ($x=0,1,2,...,D-1$), and $k$ assumes
the values $k=2\pi n/D$ ($k$ lives in the {\sl reciprocal space}
of the crystal), with integer $-(D-1)/2\leq n\leq (D-1)/2$ (The
electrons are, in this example, spinless). The filled states have
$|k| \leq 2\pi(N-1)/(2D)=k_F$ where the {\sl Fermi-wavenumber} $k_F$
is $2\pi(N-1)/(2D)\simeq\pi\nu$ for $N >>
1$. We assume that $N$ is odd. Here $\nu=\frac{N}{D}$ is the {\sl
filling} of the band (the number of electrons per orbital
$\ket{x}$).  Note that the {\sl empty states} are those with
$k_F<|k|<\pi$, and $k=\pi$, but {\em not} $k=-\pi$, corresponding
to the rule that $k$s differing by a {\sl reciprocal lattice
vector}, in particular those lying on the boundary of the {\sl
first Brillouin zone}, should not be counted twice.

Suppose that a measurement is done that reveals that an electron
is present at the origin.  What is the new Slater determinant
describing the state?  That is, we are specializing the
development in the text to the case
\begin{eqnarray}
\ket{\kappa}=\ket{x=0}={1\over\sqrt{D}}\sum_k\ket{k},\\
\ket{\mbox{in}}={1\over\sqrt{N}}\sum_{|k|\leq k_F}\ket{k}=\ket{W_0},\\
\ket{\mbox{out}}={1\over\sqrt{D-N}}\sum_{|k|> k_F}\ket{k},\\
\ket{\kappa}=\sqrt{\nu}\,\ket{\mbox{in}}+\sqrt{1-\nu}\,\ket{\mbox{out}},\\
\ket{W_s}={1\over\sqrt{N}}\sum_{|k|\leq
k_F}e^{iks}\ket{k},\,\,\,s=0,1,...,N-1.
\end{eqnarray}
Here we have introduced the orbitals $\ket{W_s}$, which are
obtained by a Fourier transform over the {\em filled states}
$\ket{k}$.  They are somewhat localized on the lattice, but not
perfectly, since they only include the plane waves up to a certain
wavelength.  The wave function of $W_0$ is
\begin{equation}
W_0(x)=\bra{x}W_0\rangle={\sin(\pi\nu x)\over\pi\sqrt{\nu}x}.
\end{equation}
The other orbitals $\ket{W_s}$, $s\neq 0$, are basically displaced
versions of $W_0$:
\begin{equation}
W_s(x)=W_0(x-s/\nu).
\end{equation}
Note, however, that an analytic continuation of $x$ to the reals
is understood here, since for general $\nu$ the $W_s$ wave
functions are generally not centered on lattice sites.

The orbitals $\ket{W_s}$ rather resemble the {\sl Wannier
functions} of band theory, in that they are approximately
localized states built out of band orbitals.  They are different,
though, in that Wannier functions are generally defined for full
bands only, i.e., only for $\nu=1$.

So, again, how is the fermi sea modified if the electron number is
measured at the origin?  With probability $\nu$ the answer will be
``1", and then the new Fermi sea has the $W_0$ orbital replaced by
the completely localized orbital $\ket{\kappa}=\ket{0}$, and all
the rest unperturbed:
\begin{equation}
a_\kappa^\dagger\Pi_{s\neq 0}a_{W_s}^\dagger\ket{\bf 0}.
\end{equation}
One can say that one orbitals' worth of electrons has been pulled
out from $O(1/\nu)$ lattice sites around the origin and
concentrated at $x=0$.  The hole that is left in the fermi sea by
this process is what is known as the {\sl exchange hole} in
electron physics \cite{book:pines}.

With probability $1-\nu$ the measurement gives answer ``0"; then
the $W_0$ orbital replaced by $\ket{\kappa^\perp}$:
\begin{equation}
a_{\kappa^{\perp}}^\dagger\Pi_{s\neq 0}a_{W_s}^\dagger\ket{\bf 0}.
\end{equation}
This modified orbital can be rewritten in an informative way:
\begin{equation}
\ket{\kappa^\perp}=\sqrt{1-\nu}\,\ket{\mbox{in}}-\sqrt{\nu}\,\ket{\mbox{out}}
=-\sqrt{\nu\over1-\nu}\ket{0}+\sqrt{1\over1-\nu}\ket{W_0}.
\end{equation}
We find that the wave function for this orbital
$\bra{x}\kappa^\perp\rangle$ is zero for $x=0$, as expected, and
also has an exchange hole, but with a reversed sign compared with
the other measurement outcome, and with a magnitude that depends
on $\nu$.  If $\nu$ is near one, the perturbation of the fermi sea
is very strong, but also this outcome ``0" occurs with vanishingly
small probability.

\bibliographystyle{hunsrt}
\bibliography{refs}

\end{document}